\documentclass[journal]{IEEEtran}
\usepackage{cite,mathtools}
\usepackage{graphicx}
\usepackage{subfigure}
\usepackage{amssymb}
\usepackage{xcolor}
\usepackage{url,breakurl}
\usepackage{float}
\usepackage{booktabs}
\usepackage{bigstrut}
\usepackage{graphicx}
\usepackage{subfigure}
\usepackage{array}
\usepackage{color}
\newcommand{\tabincell}[2]{\begin{tabular}{@{}#1@{}}#2\end{tabular}}
\usepackage{url}

\begin{document}

\title{6G Wireless Channel Measurements and Models: Trends and Challenges}
%
%
%

\author{Cheng-Xiang~Wang,~\IEEEmembership{Fellow,~IEEE}, Jie~Huang,~\IEEEmembership{Member,~IEEE},
Haiming Wang,~\IEEEmembership{Member,~IEEE},
Xiqi~Gao,~\IEEEmembership{Fellow,~IEEE},
Xiaohu You,~\IEEEmembership{Fellow,~IEEE},
and Yang Hao,~\IEEEmembership{Fellow,~IEEE}
\thanks{C.-X. Wang (corresponding author), J. Huang, X. Gao, and X. You are with the National Mobile Communications Research Laboratory, School of Information Science and Engineering, Southeast University, Nanjing, 210096, China, and also with the Purple Mountain Laboratories, Nanjing, 211111, China (e-mail: \{chxwang, j\_huang, xqgao, xhyu\}@seu.edu.cn).}
\thanks{H. Wang is with the State Key Laboratory of Millimeter Waves, School of Information Science and Engineering, Southeast University, Nanjing, 210096, China, and also with the Purple Mountain Laboratories, Nanjing, 211111, China (e-mail: hmwang@seu.edu.cn).}
\thanks{Y. Hao is with the School of Electronic Engineering and Computer Science, Queen Mary University of London, London, E1 4NS, U.K. (e-mail: y.hao@qmul.ac.uk).}
\thanks{The authors would like to acknowledge the support from the National Key R\&D Program of China under Grant 2018YFB1801101, the National Natural Science Foundation of China (NSFC) under Grants 61960206006 and 61901109, the National Postdoctoral Program for Innovative Talents under Grant BX20180062, the Frontiers Science Center for Mobile
Information Communication and Security, the High Level Innovation and Entrepreneurial Research Team Program in Jiangsu, the High Level Innovation and Entrepreneurial Talent Introduction Program in Jiangsu, the Research Fund of National Mobile Communications Research Laboratory, Southeast University, under Grant 2020B01, the Fundamental Research Funds for the Central Universities under Grant 2242020R30001, the Huawei Cooperation Project, and the EU H2020 RISE TESTBED2 project under Grant 872172.}}

%
%

\markboth{IEEE Vehicular Technology Magazine}%
{Shell \MakeLowercase{\textit{et al.}}: Bare Demo of IEEEtran.cls for Journals}
%



\maketitle

\begin{abstract}
In this article, we first present our vision on the application scenarios, performance metrics, and potential key technologies of the sixth generation (6G) wireless communication networks. Then, 6G wireless channel measurements, characteristics, and models are comprehensively surveyed for all frequency bands and all scenarios, focusing on millimeter wave (mmWave), terahertz (THz), and optical wireless communication channels under all spectrums, satellite, unmanned aerial vehicle (UAV), maritime, and underwater acoustic communication channels under global coverage scenarios, and high-speed train (HST), vehicle-to-vehicle (V2V), ultra-massive multiple-input multiple-output (MIMO), orbital angular momentum (OAM), and industry Internet of things (IoT) communication channels under full application scenarios. Future research challenges on 6G channel measurements, a general standard 6G channel model framework, channel measurements and models for intelligent reflection surface (IRS) based 6G technologies, and artificial intelligence (AI) enabled channel measurements and models are also given.
\end{abstract}


%
\IEEEpeerreviewmaketitle

\section{Introduction}
%
%
%
%
In terms of application requirements, making communications ``mobile'' and ``broadband'' was the major evolution from the first generation (1G) to the fourth generation (4G) wireless communication networks, while the fifth generation (5G) has expanded from mobile broadband in 4G to enhanced mobile broadband (eMBB) plus the Internet of things (IoT). The IoT further includes massive machine type communications (mMTC) and ultra-reliable and low latency communications (uRLLC). From 2020, 5G wireless communication networks have been deployed worldwide. However, 5G will not be able to meet all the requirements of future networks. The researches on the sixth generation (6G) wireless communication networks have therefore started, which are planned to be deployed after 2030 \cite{Zhang19}. 

While 5G mainly concentrates on eMBB, mMTC, and uRLLC, 6G wireless communication networks are expected to further enhance mobile broadband, expand the boundary and coverage of the IoT, and make the networks/devices more intelligent. In \cite{Zhang19}, the authors named the enhanced three scenarios as further-eMBB (feMBB), ultra-mMTC (umMTC), and enhanced-uRLLC (euRLLC). Several other application scenarios, such as long-distance and high-mobility communications, and extremely-low power communications, are also promising. Here we classify the application scenarios as strengthened eMBB/mMTC/uRLLC and other new scenarios. The new scenarios include space-air-ground-sea integrated networks, artificial intelligence (AI) enabled networks, etc.

Driven by new application requirements, 6G has to introduce new technical requirements and performance metrics. The peak data rate for 5G is 20 Gbps, while it can be 1-10 Tbps for 6G networks due to the use of terahertz (THz) and optical wireless bands. The user experienced data rate can achieve Gbps level with the aid of high frequency bands. The area traffic capacity can be more than 1 Gbps/m$^2$. The spectrum efficiency can increase 3-5 times, while the energy efficiency can increase about 10 times compared to 5G, by applying AI to have better network management. The connection density will increase 10--100 times due to the use of extremely heterogeneous networks (HetNets), diverse communication scenarios, and large bandwidths of high frequency bands. The mobility will be be supported to higher than 1000 km/h due to the movements of ultra-high-speed trains, unmanned aerial vehicles (UAVs), and satellites. The latency is expected to be less than 1~ms. In addition, other important performance metrics should be introduced, e.g., cost efficiency, security capacity, coverage, intelligence level, etc., to evaluate 6G networks in a more comprehensive way.

To meet the above application requirements and performance metrics, 6G communication networks will have new paradigm shifts and rely on new enabling technologies. The new paradigm shifts can be summarized as global coverage, all spectrums, full applications, and strong or endogenous security. To provide global coverage, 6G wireless communication networks will expand from terrestrial communication networks in 1G--5G to space-air-ground-sea integrated networks, including satellite, UAV, terrestrial ultra-dense networks (UDNs), underground communications, maritime communications, and underwater accoustic communications. To provide higher data rate, all spectrums will be fully explored, including sub-6 GHz, millimeter wave (mmWave), THz, and optical wireless bands. With the aid of AI and big data techniques, the key technologies and applications will be highly integrated to enable full applications. Furthermore, AI can enable dynamic orchestration of networking, caching, and computing resources to improve the network performance. The last but not the least trend for 6G is to enable strong or build-in network security when developing it, including physical layer and network layer security. This is quite different from the development strategy of 1G--5G, which first make networks work and then consider whether the networks are secure and how to improve the network security.

\begin{figure*}[tb!]
\centering
\includegraphics[width=6in]{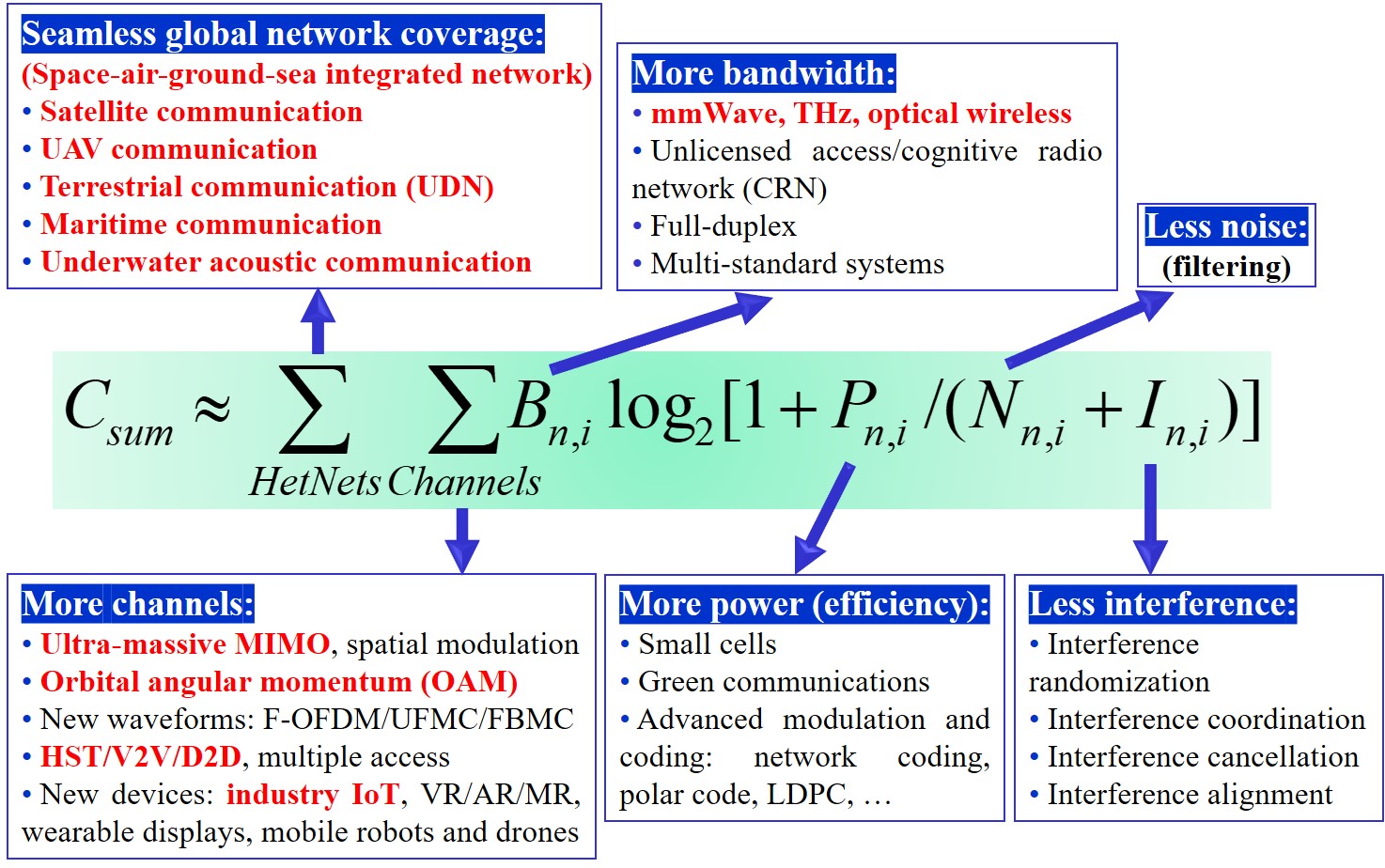}
\caption{An illustration of 6G enabling technologies.}
\label{fig:Enable_tech}
\end{figure*}

The 6G enabling technologies aim to greatly increase the sum capacity, which is approximated by the summation of Shannon link capacities of different types of channels over HetNets considering interference. As illustrated in Fig. \ref{fig:Enable_tech}, the sum capacity can be increased by increasing the signal bandwidth, signal power, number of channels in space/time/frequency domains, and number of HetNets or coverage, as well as reducing the interference and noise, thus increasing the signal-to-interference-plus-noise ratio (SINR). 

To realize 6G networks with the new trends and enabling technologies, the underlying 6G wireless channels need to be thoroughly studied, since wireless channel is the foundation for the system design, network optimization, and performance evaluation of 6G networks. In this article, 6G wireless channel measurements, characteristics, and models are comprehensively surveyed for all frequency bands and all scenarios, focusing on mmWave, THz, and optical wireless communication channels under all spectrums, satellite, UAV, maritime, and underwater accoustic communication channels under global coverage scenarios, and high-speed train (HST), vehicle-to-vehicle (V2V), ultra-massive multiple-input multiple-output (MIMO), orbital angular momentum (OAM), and industry IoT communication channels under full application scenarios. Future challenges on 6G channel measurements, a general standard 6G channel model framework, channel measurements and models for intelligent reflection surface (IRS) based 6G technologies, and AI enabled channel measurements and models are also given.

The remainder of the paper is organized as follows. In Section \ref{Sec2}, we summarize channel measurements and characteristics for different types of 6G channels. Section \ref{Sec3} studies the channel models for all frequency bands and all scenarios. In Section \ref{Sec4}, we provide some future research challenges. Conclusions are drawn in Section~\ref{Sec5}.

\section{6G Channel Measurements and Characteristics}\label{Sec2}
\begin{figure*}[tb!]
\centering
\includegraphics[width=5.5in]{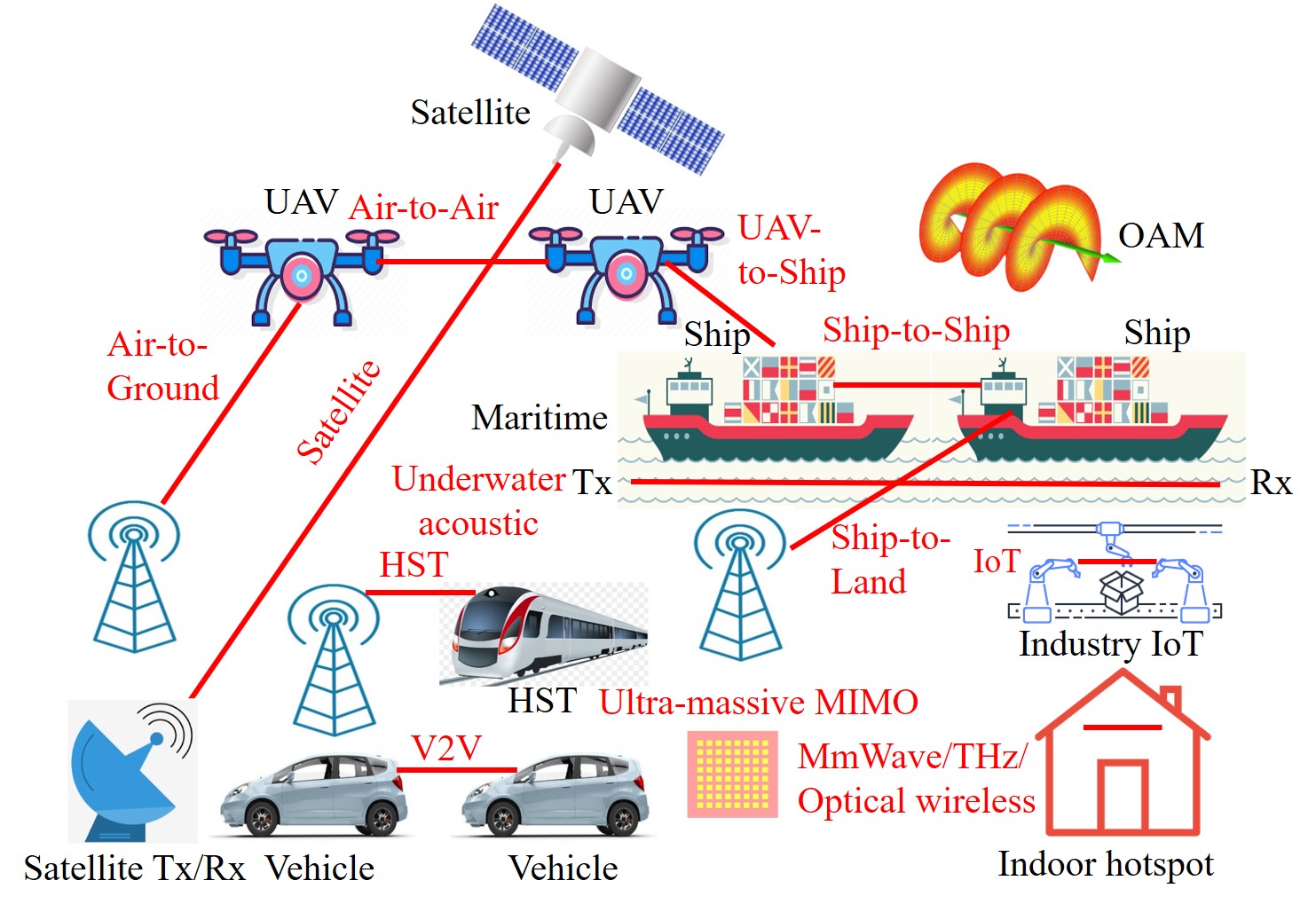}
\caption{An illustration of different types of 6G wireless channels.}
\label{fig:6G_channel}
\end{figure*}

6G wireless channels are existed at multiple frequency bands and in multiple scenarios, as illustrated in Fig. \ref{fig:6G_channel}. The channel sounders and channel characteristics for each individual channel show great differences \cite{Wang18}. Here a comprehensive survey of different types of 6G channels is presented by grouping them under all spectrums, global coverage scenarios, and full application scenarios. A summary of 6G channel measurements and characteristics is shown in Table 1.

\begin{table*}[tb!]
\caption{A summary of 6G channel measurements and characteristics.}
\label{tab:measurement}
\centering
  \arraybackslash
\begin{tabular}{|m{2.4cm}<{\centering}|m{2.8cm}<{\centering}|m{3.2cm}<{\centering}|m{6.5cm}<{\centering}|}
\hline
\textbf{Wireless channel}&\textbf{Measured frequency bands}&\textbf{Measured scenarios}&\textbf{Channel characteristics}\\
\hline
MmWave/THz channel&26/28, 32, 38/39,
60, and 73 GHz bands (mmWave); around 300 GHz (THz)& Indoor and outdoor&Large bandwidth, high directivity, large path loss, blockage effects, atmosphere absorption, more diffuse scattering\\
\hline
Optical wireless channel&Mainly 380-780 nm& Indoor, outdoor, underground, underwater&Complex scattering properties for different materials, non-linear photoelectric characteristics at Tx/Rx ends, background noise effects\\
\hline
Satellite channel&Ku, K, Ka, and V bands& GEO, LEO, MEO, and HEO&Rain/cloud/fog/snow attenuation, extremely large Doppler frequency shift and Doppler spread, frequency dependence, large coverage range, long communication distance\\
\hline
UAV channel&2, 2.4, and 5.8 GHz& Urban, suburban, rural, and open field (air-to-air and air-to-ground)&3D random trajectory (large elevation angle), high mobility, spatial and temporal non-stationarity, airframe shadowing\\
\hline
Maritime channel&2.4 and 5.8 GHz& UAV-to-ship, ship-to-ship, and ship-to-land&Sparse scattering, sea wave movement, ducting effect over the sea surface, time non-stationary, long communication distances, climate factors\\
\hline
Underwater acoustic channel&2-32 kHz& Underwater environments&High transmission loss, multipath propagation, time-varying, Doppler effects\\
\hline
HST/V2V channel&Sub-6 GHz and mmWave bands& Open space, hilly terrain, viaduct, tunnels, cutting, stations, and intra-wagon (HST); highway, urban street, open area, university campus, and parking lot (V2V)&Large Doppler frequency shift and Doppler spread, non-stationarity, effect of train/vehicle, velocity and trajectory variations\\
\hline
Ultra-massive MIMO channel&Sub-6 GHz, mmWave, and THz bands& Indoor and outdoor&Spatial non-stationarity, channel hardening, spherical wavefront\\
\hline
OAM channel&mmWave& LOS and NLOS (reflection)&Multiplexing gain, beam divergence and misalignment, degradation in reflection scenarios\\
\hline
Industry IoT channel&Sub-6 GHz& Industry IoT environments&Varied path loss, random fluctuations, NLOS propagation, large amounts of scatterers, multi-mobility\\
\hline
\end{tabular}
\end{table*}

\subsection{6G channel measurements and characteristics for all spectrums}

\subsubsection{MmWave/THz channel}
In general, mmWave refers to 30--300 GHz band, while THz denotes 0.1--10 THz. Thus, the 100--300~GHz band shares some common characteristics with mmWave and THz, such as large bandwidth, high directivity, large path loss, blockage effects, atmosphere absorption, and more diffuse scattering \cite{Huang19, Rap19, Huq19}. While mmWave is applied to achieve Gbps level transmission data rate up to several hundred meters with several GHz bandwidths, THz is known to achieve Tbps level transmission data rate up to tens of meters with several tens of GHz bandwidths. THz bands show more severe path loss, atmosphere absorption, and diffuse scattering than mmWave bands.

\begin{figure*}[bt!]
\centering
\begin{minipage}[t]{0.45\linewidth}
\centerline{\includegraphics[width=3.2in]{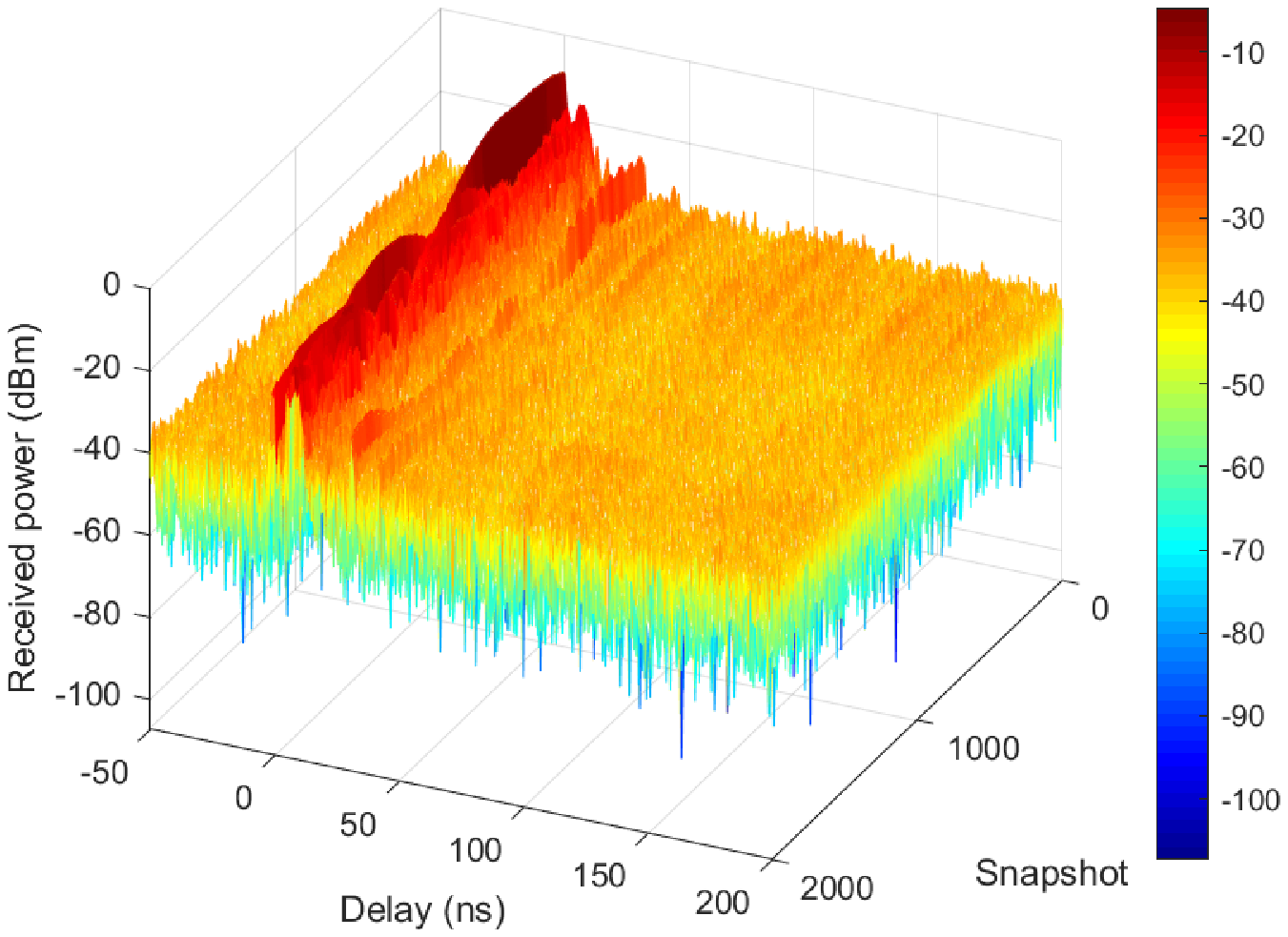}}
\footnotesize \centerline{(a) Measured mmWave V2V channel at 28 GHz band.}
\end{minipage}
\begin{minipage}[t]{0.45\linewidth}
\centerline{\includegraphics[width=3.2in]{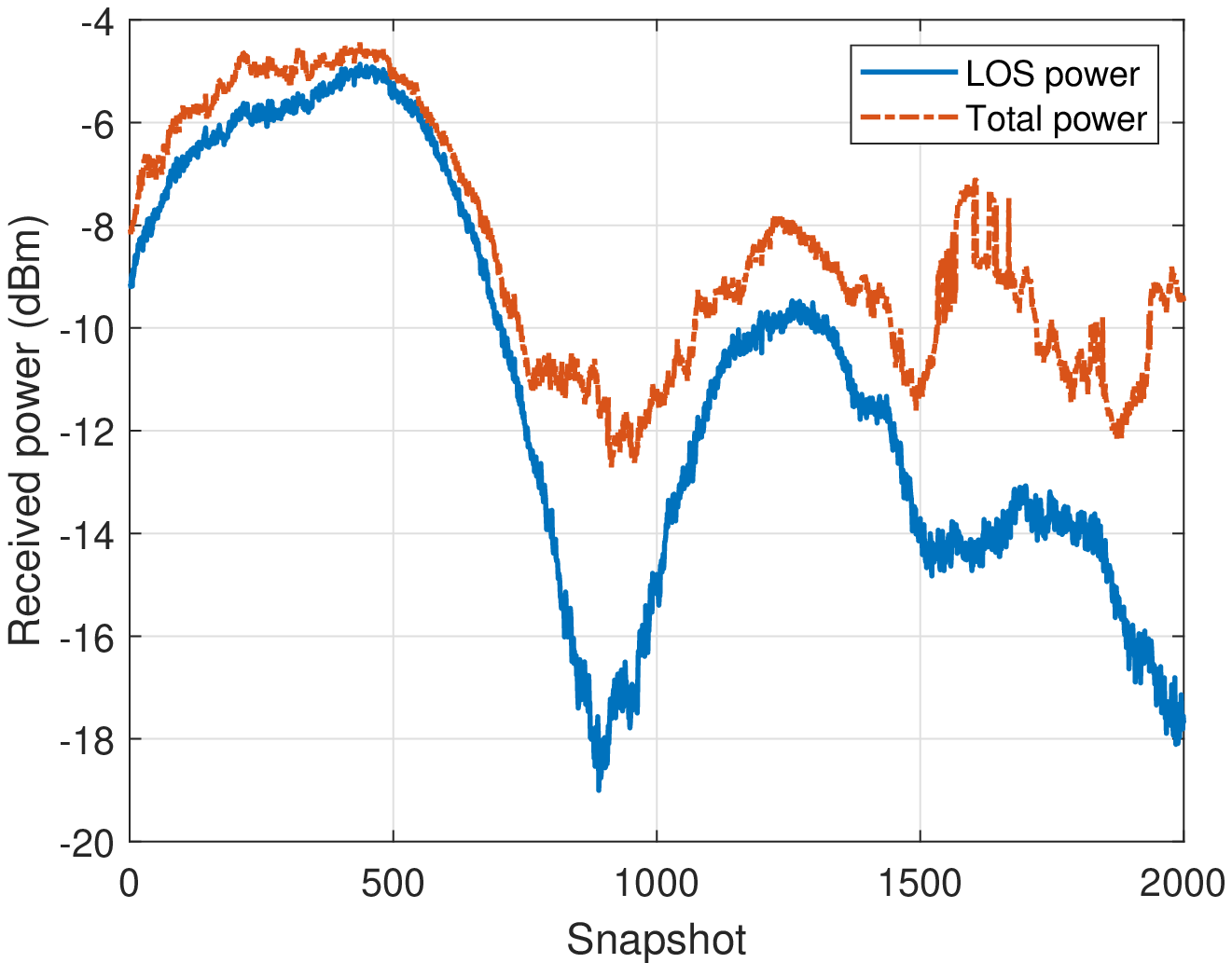}}
\footnotesize \centerline{(b) Received power variations.}
\end{minipage}
\caption{The measured mmWave V2V channel variations at 28 GHz band.}
\label{fig:mmWave_cm}
\end{figure*}

MmWave channel has been well studied at some typical frequency bands, such as 26/28, 32, 38/39, 60, and 73 GHz bands. Even though, mmWave channel measurements with MIMO antennas, high dynamics (such as V2V), and outdoor environments are still needed. An illustration of the measured 28 GHz mmWave V2V channel is shown in Fig. \ref{fig:mmWave_cm}, which is obtained from our real channel measurements. The line-of-sight (LOS) power and total power variate over the 2000 snapshots, which validates the non-stationarity of the channel. In \cite{Huang19}, the recent developments and future challenges on mmWave channel sounders and measurements were given. In \cite{Rap19}, some preliminary path loss, partition loss, and scattering measurements were conducted at 140 GHz. Most of the current THz channel measurements are around 300 GHz band. The channel characteristics above 300 GHz are still not clear, which need extensive channel measurements in the future.

\subsubsection{Optical wireless channel}
Optical wireless bands refer to electromagnetic spectrums with carrier frequencies of infrared, visible light, and ultraviolet, which corresponds to wavelengths in the range of 780--10$^6$ nm, 380--780 nm, and 10--380 nm, respectively \cite{Kin18}. They can be used for wireless communications in indoor, outdoor, underground, and underwater scenarios. Optical wireless channel shows some unique channel characteristics, such as complex scattering properties for different materials, non-linear photoelectric characteristics at transmitter/receiver (Tx/Rx) ends, background noise effects, etc. The channel scenarios can be further classified as directed LOS, non-directed LOS, non-directed non-LOS (NLOS), tracked, etc. \cite{Kin18}. The main differences between optical wireless and traditional frequency bands are that there are no multipath fading, Doppler effect, and bandwidth regulation. The measured channel parameters include channel impulse response/channel transfer function (CIR/CTF), path loss, shadowing fading, root mean square (RMS) delay spread, etc.

\subsection{6G channel measurements and characteristics for global coverage scenarios}

\subsubsection{Satellite channel}
Satellite communication has attracted much interest in current wireless communication systems and is deemed to provide global coverage due to its feasible services and lower cost \cite{Sae17}. In general, satellite communication orbits can be divided as geosynchronous orbit and non-geostationary orbit. The circular geosynchronous Earth orbit (GEO) is 35786 km above Earth's equator and follows the direction of Earth's rotation. Non-geostationary orbits can be further classified as low Earth orbit (LEO), medium Earth orbit (MEO), and high Earth orbit (HEO), depending on the distance of satellites to the Earth. The usually applied frequency bands are Ku (12-18 GHz), K (18-26.5 GHz), Ka (26.5-40 GHz), and V (40-75 GHz) bands. The satellite communication channel is largely affected by weather dynamics, including rain, cloud, fog, snow, etc. Rain is the major source of attenuation, especially at frequency bands above 10 GHz. Besides, satellite communication channel shows extremely large Doppler frequency shift and Doppler spread, frequency dependence, large coverage range, long communication distance, etc. As the distance is extremely long, the channel can be viewed as LOS transmission and multipath effects can be ignored. Meanwhile, high transmitted power and high antenna gains are needed to combat the high path loss caused by the long distance and high frequency bands.
 
\subsubsection{UAV channel}
UAV has boosted in recent years for both civil and military applications. The UAV channel shows some unique channel characteristics, such as three-dimensional (3D) deployment, high mobility, spatial and temporal non-stationarity, and airframe shadowing \cite{Khu18, Kha19}. In general, UAV channel can be classified as air-to-air and air-to-ground channels. Two types of aerial vehicles are used for channel measurements, i.e., small/medium sized manned aircraft and UAVs. Channel measurements for the first kind is expensive, while the second kind can largely reduce the cost \cite{Khu18}. Both narrowband and wideband channel measurements have been conducted, most of which are at 2, 2.4, and 5.8 GHz bands. The measured environments include urban, suburban, rural, and open field. The measured channel parameters include path loss, shadowing fading, RMS delay spread, K-factor, amplitude probability density function (PDF)/cumulative distribution function (CDF), etc.

\subsubsection{Maritime channel}
As a part of the space-air-ground-sea integrated networks, maritime communication channel mainly includes air-to-sea and near-sea-surface channels \cite{Wang18_2}. For air-to-sea channel, the UAV or relay is used as the base station (BS) to communicate with ships on the sea surface. This type of channel is also named as UAV-to-ship channel. For near-sea-surface channel, a ship can communicate with other ships (ship-to-ship) or fixed BS near the sea (ship-to-land). The unique features of maritime propagation environment causes many new channel characteristics, such as sparse scattering, sea wave movement, ducting effect over the sea surface, time non-stationary, long communication distances, and climate factors, which show great differences from conventional terrestrial wireless channels. Maritime channel measurements were conducted at 2.4 GHz and 5.8 GHz bands with maximum distances up to 10 km \cite{Wang18_2}. The path loss, RMS delay spread, and K-factor were studied.

\subsubsection{Underwater acoustic channel}
The underwater acoustic channel faces many challenges. Because of the ambient noise in the oceans, the applicable frequency is low and the transmission loss is high. Horizontal underwater channels are prone to multipath propagation due to refraction, reflection, and scattering. The underwater acoustic channel disperses in both time and frequency domain, which leads to the time-varying and Doppler effects. Channel
measurements were unusually conducted at several kHz, ranging from 2 kHz to 32 kHz.

\subsection{6G channel measurements and characteristics for full application scenarios}

\subsubsection{HST/V2V channel}
The previous HST communication systems are mainly global system for mobile communication railway (GSM-R) and long term
evolution for railway (LTE-R). Recently, 5G network is being applied to HST to improve the quality of services (QoS) \cite{Liu19}. The speed of ultra-HST is desired to exceed 500 km/h in the future, which causes problems such as frequent and fast handover and large Doppler spread. MmWave/THz and massive MIMO are potential key technologies to be utilized in HST communication systems. Some preliminary channel measurements have been conducted for HST environments, including open space, hilly terrain, viaduct, tunnels, cutting, stations, and intra-wagon \cite{Liu19}.

Vehicular network is a typical industry vertical application of 5G/6G for uRLLC scenario. The channels include V2V, vehicle-to-infrastructure (V2I), vehicle-to-pedestrian (V2P), and are called as vehicle-to-everything (V2X) in general. V2V channel at sub-6 GHz band has been widely investigated, while mmWave V2V channel needs more measurements. A survey of current mmWave V2V channel measurements was given in \cite{He19}. In summary, V2V channels were measured at 28, 38, 60, 73, and 77 GHz bands. All of them are configured with single antenna at both sides. The measured environments include highway, urban street, open area, university campus, parking lot, etc. MmWave V2V MIMO or even massive MIMO channel measurements with high mobility are promising in the future. How to measure it in a efficient and low-cost way is still an open issue. 

\subsubsection{Ultra-massive MIMO channel}
Ultra-massive MIMO utilizes thousands of antennas to largely improve the spectral and energy efficiency, throughput, robustness, and degree of freedoms of wireless communication systems. It can be combined with other key technologies, such as mmWave/THz, V2V, and HST communications. Due to the use of large antenna array, the channel shows spherical wavefront, spatial non-stationarity, and channel hardening properties, which have been validated by previous massive MIMO channel measurements at sub-6 GHz/mmWave bands in indoor and outdoor environments. At sub-6 GHz band, the dimension of the massive MIMO array can be several meters. At THz band, due to the developments of plasmonic nano-antenna arrays, it is possible to realize ultra-massive MIMO up to 1024$\times$1024 \cite{Aky16}. For 0.06--1~THz band, metamaterials enable the design of plasmonic nano-antenna arrays with hundreds of elements in a few square centimeters. For 1--10 THz band, graphene-based plasmonic nano-antenna arrays with thousands of elements can be embedded in a few square millimeters \cite{Aky16}.

\subsubsection{OAM channel}
OAM has attracted a widespread interest in many fields, especially in telecommunications due to its potential to increase capacity by multiplexing. The number of orthogonal OAM modes in a single beam is theoretically infinite and each mode is an element of a complete orthogonal basis that can be employed for multiplexing different signals, thus greatly improving the spectrum efficiency. OAM represents electron rotation around the propagation axis generated by the energy flow. OAM based communication can be obtained from traditional MIMO theory under certain conditions. However, beam divergence and misalignment will severely decrease the transmission distance of OAM waves. Moveover, reflection will destroy orthogonality of OAM waves, thus degrading the performance in NLOS scenario. Up to now, there are limited channel measurements to verify the feasibility of OAM in different scenarios.

\subsubsection{Industry IoT channel}
In industry IoT environments, there are various robots, sensors, and mechanical devices which need massive connections in a robust and efficient manner \cite{Wang19}. The industry IoT channel exhibits many new channel characteristics, such as varied path loss, random fluctuations, NLOS propagation, large amounts of scatterers, and multi-mobility. Only a few channel measurements have been conducted in industry IoT environments, which are mainly at sub-6 GHz band as in current IoT standards. However, channel measurements at mmWave bands are also promising in industry IoT environments for future massive connections with high transmission data rate.

\section{6G Channel Models for All Frequency Bands and All Scenarios}\label{Sec3}

Large-scale channel characteristics consist of path loss and shadowing fading, while small-scale channel characteristic is caused by multipath fading. In general, channel models can be classified as deterministic and stochastic models. Deterministic channel models include measurement-based model and ray tracing model. The map-based model and point-cloud model are simplified ray tracing models. The stochastic models can be further classified as geometry based stochastic model (GBSM), correlation based stochastic model (CBSM), and beam domain channel model (BDCM). Deterministic channel models are suitable for link-level simulation and can achieve high accuracy at the cost of high computing complexity, while stochastic channel models are the trade-off of acceptable accuracy, moderate complexity, and adaptable flexibility, thus are suitable for system-level simulation. GBSM includes pure-GBSM and semi-GBSM. Pure-GBSM can be classified as regular-shaped and irregular-shaped ones. Semi-GBSM is adopted in many standardized channel models. Due to the unique channel characteristics of different type of 6G wireless channels, many large-scale and small-scale channel models have been proposed by using different channel modeling methods to accurately describe the underlying channels.

\subsection{6G channel models for all spectrums}

\begin{figure*}[bt!]
\centering
\begin{minipage}[t]{0.45\linewidth}
\centerline{\includegraphics[width=3.2in]{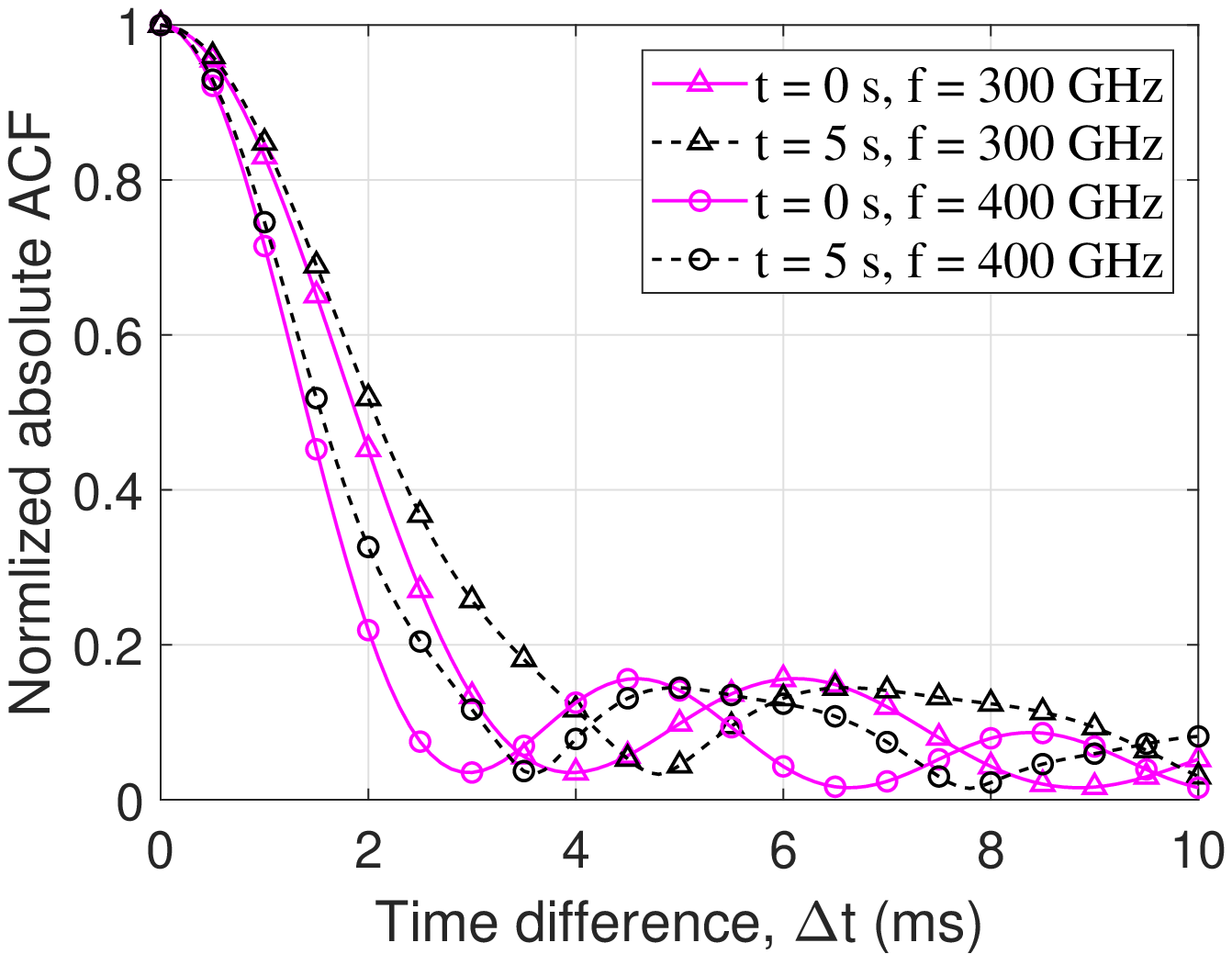}}
\footnotesize \centerline{(a) Temporal ACF for THz channel.}
\end{minipage}
\begin{minipage}[t]{0.45\linewidth}
\centerline{\includegraphics[width=3.2in]{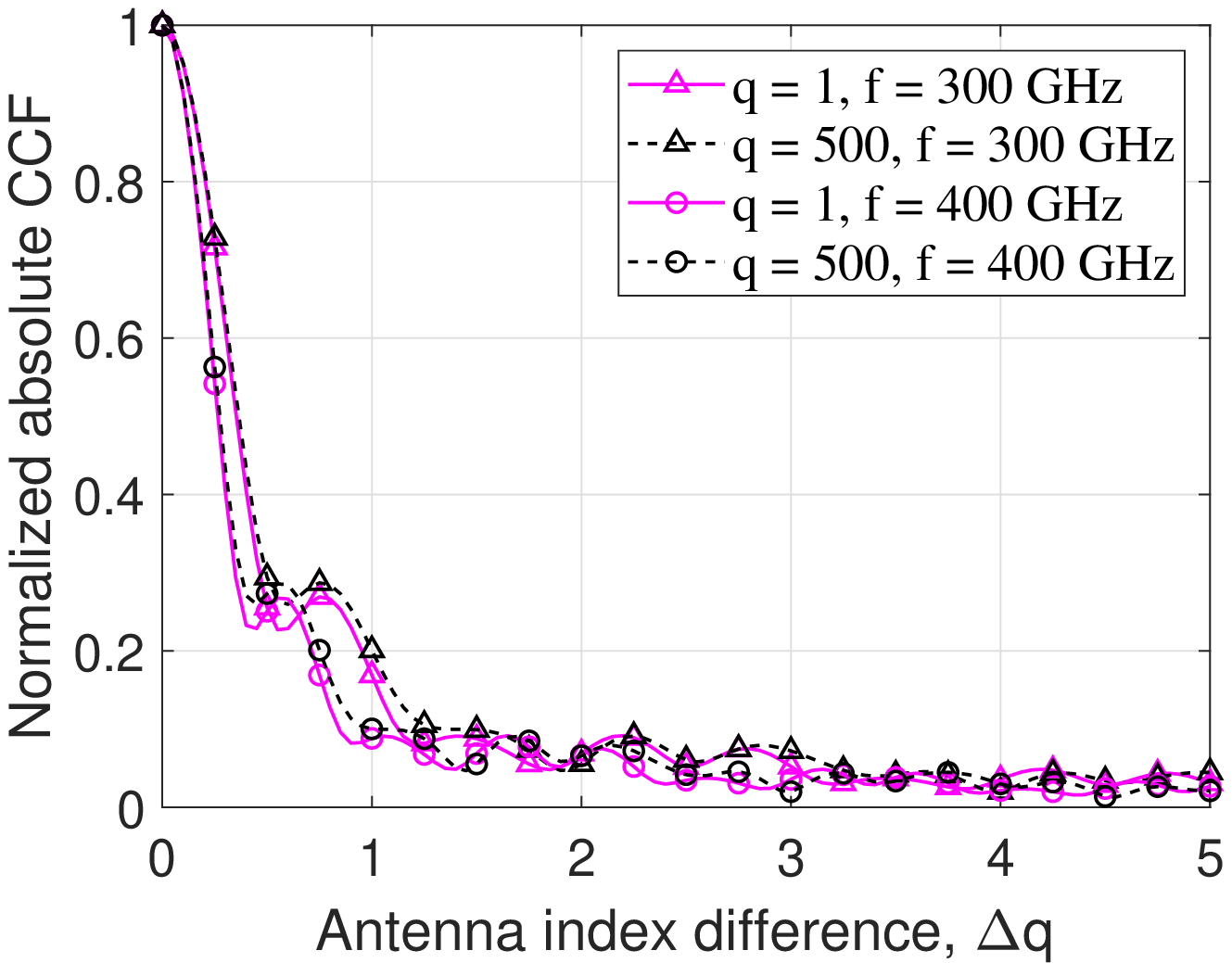}}
\footnotesize \centerline{(b) Spatial CCF for THz channel.}
\end{minipage}
\caption{The temporal ACF and spatial CCF for THz channel.}
\label{fig:THz_cm}
\end{figure*}

In \cite{Huang19}, mmWave channel models were surveyed. The deterministic channel models include ray tracing, map-based, and point cloud models. The ray tracing model is applied to IEEE 802.11ad, while the map-based model is applied to METIS. The quasi-deterministic (Q-D) model is used in MiWEBA and IEEE 802.11ay. The stochastic models include SV, propagation graph, and GBSM. GBSM is used in several standardized channel models, such as NYU WIRELESS, 3GPP 38.901, METIS, and mmMAGIC. The ray tracing model and GBSM are also widely used in THz channel modeling. Meanwhile, human/vegetation blockage, rain/cloud/snow/fog attenuations are also need to be modeled for mmWave/THz channel. The temporal autocorrelation function (ACF) and spatial cross-correlation function (CCF) for THz channel are presented in Fig. \ref{fig:THz_cm}. As the frequency increases, the temporal ACF and spatial CCF tend to be smaller with the same time difference and antenna index difference.

For optical wireless channel, the proposed deterministic models include recursive model, iterative model, DUSTIN model, ceiling bounce model, and geometry based deterministic model. The proposed stochastic models are classified as GBSM and non-GBSM. A detailed description of each optical wireless channel model was given in \cite{Kin18}.

\subsection{6G channel models for global coverage scenarios}

As satellite communication channel is mainly LOS transmission, the received signal in stable in general, except the effects of weather condition and tropospheric scintillation. Most of current channel models are concerned about the PDF of the received signal amplitude. According to the received signal strength, the channel condition can be classified as good, moderate, and bad, which can be modeled by using Markov-chain. Meanwhile, some preliminary works try to use GBSM to model the satellite channel.

A comprehensive summary of air-to-ground large-scale path models was given in \cite{Khu18}. UAV small-scale channel models include deterministic and stochastic ones. The deterministic models include ray tracing and analytical models such as two-ray model. The stochastic models include RS-GBSM, IS-GBSM, non-GBSM, and Markov model.

Ray tracing can be used as a deterministic simulation method for maritime channel and underwater acoustic channel. Apart from it, the two-ray model and three-way model are also used in practice. Stochastic models include GBSM and two wave with diffusion power (TWDP). Rayleigh, Ricean, and log-normal distributions are usually used for underwater acoustic channel.

\subsection{6G channel models for full application scenarios}

\begin{figure}[tb!]
\centering
\includegraphics[width=3.6in]{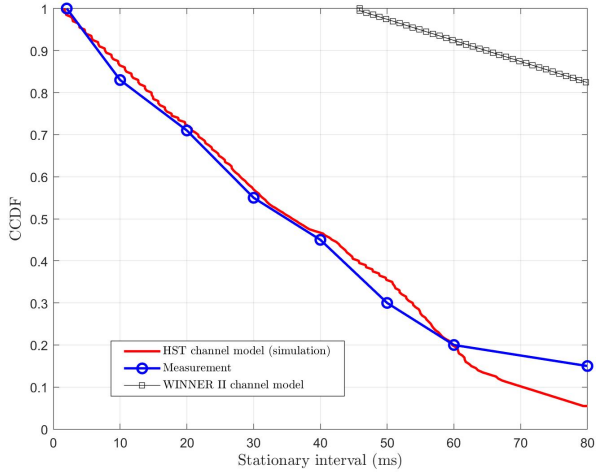}
\caption{Comparison of stationary intervals from HST channel measurements and models \cite{Wang18_3}.}
\label{fig:HST_cm}
\end{figure}

For HST and V2V channels, the high mobility and non-stationarity need to be considered. A summary of HST channel models was presented in \cite{Liu19}. Ray tracing can be used to simulate the HST/V2V channel.   Stochastic channel models include GBSM, QuaDRiGa-based model, dynamic model, Markov model, and propagation graph model. A comparison of the complementary cumulative distribution function (CCDF) of the stationary intervals from HST channel measurements and models is shown in Fig. \ref{fig:HST_cm} \cite{Wang18_3}. The proposed  general 3D non-stationary 5G channel model in \cite{Wang18_3} is more realistic than WINNER II channel model.

For ultra-massive MIMO channel, the spherical wavefront, non-stationarity, and cluster appearance and disappearance properties need to be considered. In general, the spherical wavefront can be modeled in GBSM with accurate propagation distance calculation for each individual antenna element. The non-stationrity is usually modeled by the concept of visible region and the cluster birth-death process.    

For OAM channel, the current researches focus on OAM wave generation/detection, antenna design, and the discussion of OAM potentials in wireless communications. The limited OAM channel analysis results mainly aim to verify the feasibility of OAM in different scenarios. Channel modeling for OAM wave propagation is still an open issue. 

In \cite{Wang19}, different path loss channel models were compared for industry IoT channels, including the free space path loss model, single-slope model, 3GPP models (RMa, UMa, UMi, InH), industry indoor model, and overall path loss model. The free space path loss model is used as a baseline. The single-slope model uses the apparent transmit power and path loss exponent to describe the signal strength. 3GPP models use different models for the four scenarios. The industry indoor path loss model is based on extensive channel measurement results. The overall path loss model takes LOS/NLOS condition into account to better describe the fluctuated channel status.

\subsection{Comparison of channel modeling methods for different frequency bands and scenarios}

A summary of small-scale channel models for different frequency bands and scenarios is presented in Table \ref{tab:model}. In principle, ray tracing can be used to model most types of 6G channels. However, its application to higher THz and optical wireless frequency bands needs further investigation, as the material properties at these frequency bands are lacking. It is also not applicable for satellite communication channel due to the long distance and wide area. GBSM has the widest generality and acceptable accuracy and complexity, which can be a good basis of future 6G standard channel models by assuming different geometry shapes and adding unique channel characteristics for different frequency bands and scenarios. Other modeling approaches can also provide valuable insights for the specific frequency bands and/or scenarios, such as BDCM, which converts the underlying channel to angle/beam domain. OAM channels and industry IoT channels need further study. Moreover, channel models for the combination of different frequency bands and scenarios, such as mmWave/THz + massive MIMO + HST/V2V, mmWave + satellite/UAV/industry IoT, and mmWave + maritime + UAV are challenging and need more attention in the future. 

\begin{table*}[tb!]
\caption{A summary of small-scale channel models for different frequency bands and scenarios.}
\label{tab:model}
\centering
  \arraybackslash
\begin{tabular}{|m{4.5cm}<{\centering}|m{9cm}<{\centering}|}
\hline
\textbf{Wireless channels}&\textbf{Channel models}\\ \hline
MmWave/THz channels&\tabincell{c}{\textbf{Deterministic}: ray tracing, map-based, point cloud;\\ \textbf{Stochastic}: GBSM and non-GBSM (Q-D, propagation graph)} \\ \hline
Optical wireless channels&\tabincell{c}{\textbf{Deterministic}: recursive model, iterative model, DUSTIN model, \\ceiling bounce model, and geometry based deterministic model;\\ \textbf{Stochastic}: GBSM and non-GBSM} \\ \hline
Satellite channels& \textbf{Stochastic}: GBSM and non-GBSM (Markov model) \\ \hline
UAV channels&\tabincell{c}{\textbf{Deterministic}: ray tracing, analytical models;\\ \textbf{Stochastic}: GBSM and non-GBSM (Markov model)} \\ \hline
Maritime channels&\tabincell{c}{\textbf{Deterministic}: ray tracing, two-ray model, three-ray model;\\ \textbf{Stochastic}: GBSM and non-GBSM (TWDP)} \\ \hline
Underwater acoustic channels&\tabincell{c}{\textbf{Deterministic}: ray tracing;\\ \textbf{Stochastic}: GBSM} \\ \hline
HST/V2V channels&\tabincell{c}{\textbf{Deterministic}: ray tracing;\\\textbf{Stochastic}: GBSM and non-GBSM (Markov model, \\propagation graph model)} \\ \hline
Ultra-massive MIMO channels&\tabincell{c}{\textbf{Deterministic}: ray tracing;\\\textbf{Stochastic}: GBSM and non-GBSM (BDCM, CBSM)} \\ \hline
OAM channels&Not available \\ \hline
Industry IoT channels&\tabincell{c}{\textbf{Deterministic}: ray tracing;\\\textbf{Stochastic}: GBSM} \\ \hline
\end{tabular}
\end{table*}

\section{Future Research Challenges}\label{Sec4}

\subsection{6G channel measurements}
High-performance channel sounders are important to measure 6G channels in a fast and efficient way. The mmWave channel sounders include vector network analyzer (VNA) based sounders, Keysight/NI/R\&S commercial off-the-shelf (COTS) sounders, and custom-designed sounders such as the sounders from Durham University, NYU WIRELESS, University of Southern California, National Institute of Standards and Technology (NIST), etc. \cite{Huang19}. For THz channels, most of the channel sounders are based on VNA with additional up- and down-converters to achieve different THz bands. Instead, photon modulator and detector are used for optical wireless communication channels. Other equipments/conditions, such as weather stations, UAVs, boats, waterproof materials, vehicles, and large antenna arrays are needed for specific channel measurements. Thus, 6G channel measurements are more challenging, yet it is indispensable and urgent, especially for high frequency bands, high mobility, long distance, and more severe environments. 

\subsection{A general standard 6G channel model framework}
In 5G and previous generations, the standardized channel models prefer to use a general channel model framework with different parameter sets for different scenarios. A general 3D non-stationary 5G channel model was proposed in \cite{Wang18_3} to cover the four challenging scenarios, i.e., massive MIMO, HST, V2V, and mmWave. All of the channel models are only concentrated on terrestrial communication networks and frequencies up to mmWave bands. However, 6G channels are existed over the space-air-ground-sea integrated networks and frequencies up to optical wireless bands, which will be more challenging to derive a general channel model framework. As 6G wireless channels become heterogeneous and show different scales over the wavelengths, how to describe 6G wireless channels with a general standard channel model framework is an open issue which needs careful investigations. For example, how to integrate the channel characteristics of radio frequency bands (up to THz) and optical wireless bands, terrestrial scenarios and space-air-sea scenarios, various two-dimensional (2D) and 3D mobility requirements with trajectory and speed changes? How to find the extremely complicated relationship among 6G channel characteristics, frequency bands, scenarios, and system setup parameters? How to evaluate the performance of 6G channle models in terms of accuracy, complexity, and generality?

\subsection{Channel measurements and models for IRS based 6G technologies}
IRS is a recently proposed concept beyond massive MIMO where future man-made structures are electronically active with integrated electronics and wireless communication making the entire environment “intelligent”. IRS can be implemented with ultra-massive antenna arrays and controlled by reconfigurable processing networks with the aid of AI and machine learning (ML). As the wireless channel becomes intelligent and reconfigurable, IRS shows great potentials to satisfy the future demands. Channel measurements and modeling are indispensable to validate IRS, which are open issues in the current research works.

\subsection{AI enabled channel measurements and models}
\begin{figure*}[tb!]
\centering
\includegraphics[width=5.5in]{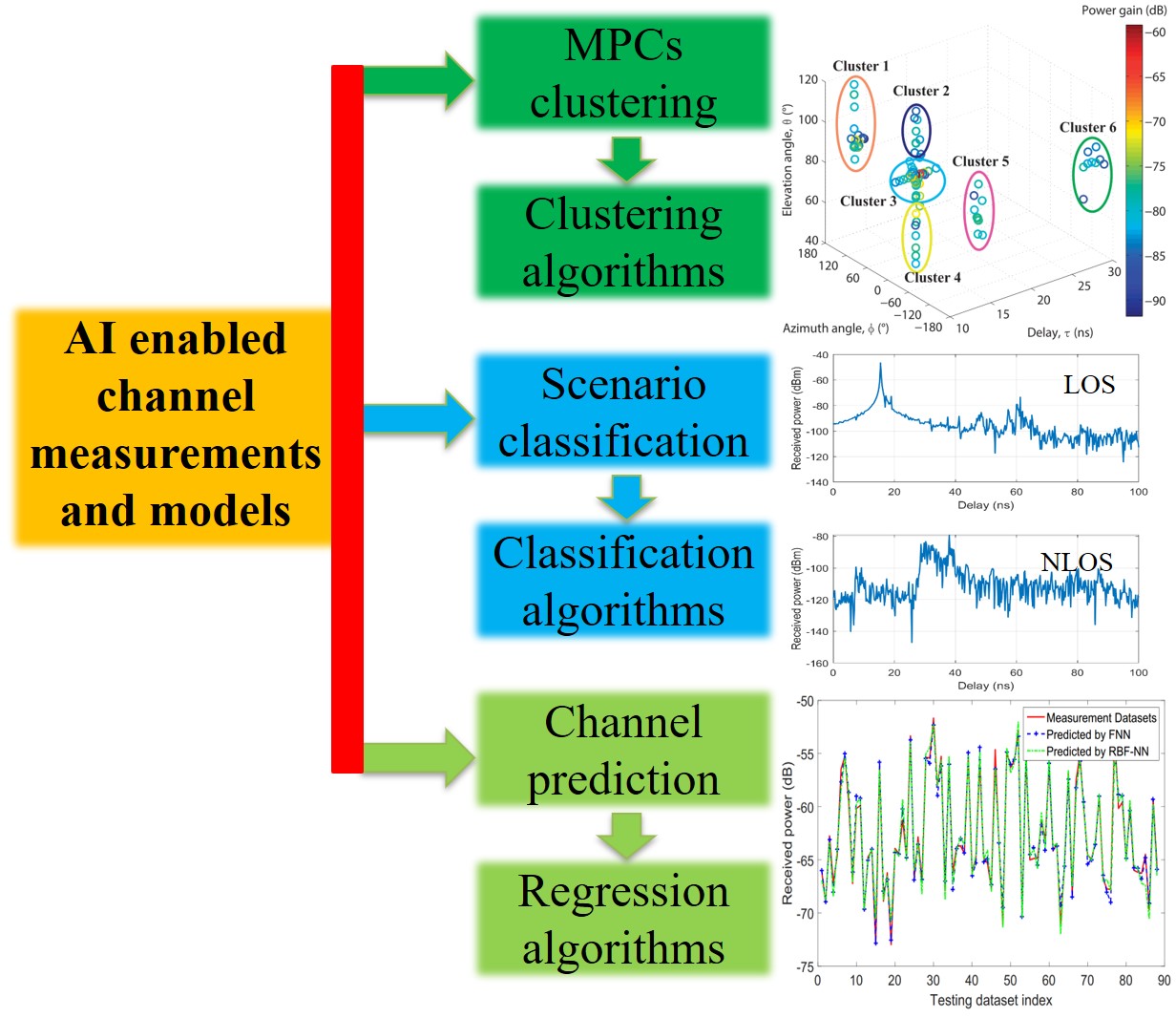}
\caption{An illustration of AI enabled channel measurements and models.}
\label{fig:AI_cm}
\end{figure*}

As the new frequency bands, scenarios, and number of antennas increase, the size of the measurement data grows rapidly, which will be too huge to process with traditional data processing methods. Some preliminary works have shown the potential of AI and ML to enable wireless channel measurements and models, for example, multipath components (MPCs) clustering, scenario classification, and channel prediction, by using clustering, classification, and regression algorithms. An illustration of AI enabled channel measurements and models is shown in Fig. \ref{fig:AI_cm}. Different ML algorithms, such as artificial neural network (ANN), convolutional neural network (CNN), and generative adversarial network (GAN) can be applied to wireless channel modeling \cite{Huang19, Liu19}. One of the benefits of applying AI and ML over traditional channel modeling methods is that they can predict wireless channel properties.

\section{Conclusions}\label{Sec5}
In this article, a vision on the new paradigm shifts of 6G wireless communication networks has been presented, as well as the performance metrics and application scenarios. A comprehensive survey of 6G channel measurements, characteristics, and models have been given to address the trends for all frequency bands and all scenarios, including mmWave, THz, and optical wireless communication channels under all spectrums, satellite, UAV, maritime, and underwater acoustic communication channels under global coverage scenarios, and HST, V2V, ultra-massive MIMO, OAM, and industry IoT communication channels under full application scenarios. More channel measurements need to be conducted for the emerging frequency bands and scenarios. In general, ray tracing and GBSM can be served as the common  deterministic and stochastic modeling methods, respectively, for most of the 6G channels by considering the individual channel characteristics. The future challenges on 6G channel measurements and models have also been pointed out.
 
%


%

\appendices






%

\begin{IEEEbiography}[{\includegraphics[width=1in,height=1.25in,clip,keepaspectratio]{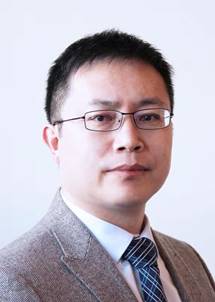}}]{Cheng-Xiang Wang} (chxwang@seu.edu.cn) received his Ph.D. degree from Aalborg University, Denmark, in 2004. He has been with Heriot-Watt University, Edinburgh, UK, since 2005, and became a professor in 2011. In 2018, he joined Southeast University, China, and Purple Mountain Laboratories, China, as a professor. His current research interests include wireless channel measurements/modeling and B5G wireless communication networks. He has co-authored four books, two book chapters, and over 390 papers in refereed journals and conference proceedings. He is a Member of Academia Europaea, a Fellow of the IEEE and IET, and a Highly Cited Researcher recognized by Clarivate Analytics in 2017--2019. 
\end{IEEEbiography}

\begin{IEEEbiography}[{\includegraphics[width=1in,height=1.25in,clip,keepaspectratio]{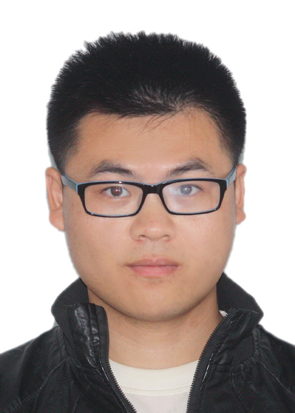}}]{Jie Huang} (j\_huang@seu.edu.cn) received the B.E. degree in Information Engineering from Xidian University, China, in 2013, and the Ph.D. degree in Communication and Information Systems from Shandong University, China, in 2018. From Jan. 2019 to Feb. 2020, he was a Postdoctoral Research Associate in Durham University, U.K. He is currently a Postdoctoral Research Associate in the National Mobile Communications Research Laboratory, Southeast University, China and also a researcher in Purple Mountain Laboratories, China. His research interests include millimeter wave and massive MIMO channel measurements and channel modeling, wireless big data, and B5G/6G wireless communications. 
\end{IEEEbiography}

\begin{IEEEbiography}[{\includegraphics[width=1in,height=1.25in,clip,keepaspectratio]{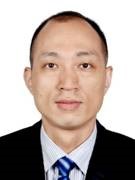}}]{Haiming Wang} (hmwang@seu.edu.cn) was born in 1975. He received the B.S., M.S., and Ph.D. degrees in Electrical Engineering from Southeast University, Nanjing, China, in 1999, 2002, and 2009, respectively. He joined the State Key Laboratory of Millimeter Waves, Southeast University, in April 2002. Now he is a professor. His current research interests include antennas and propagation for wireless communications. He was awarded for contributing to the development of IEEE 802.11aj by the IEEE-SA in July 2018.
\end{IEEEbiography}

\begin{IEEEbiography}[{\includegraphics[width=1in,height=1.25in,clip,keepaspectratio]{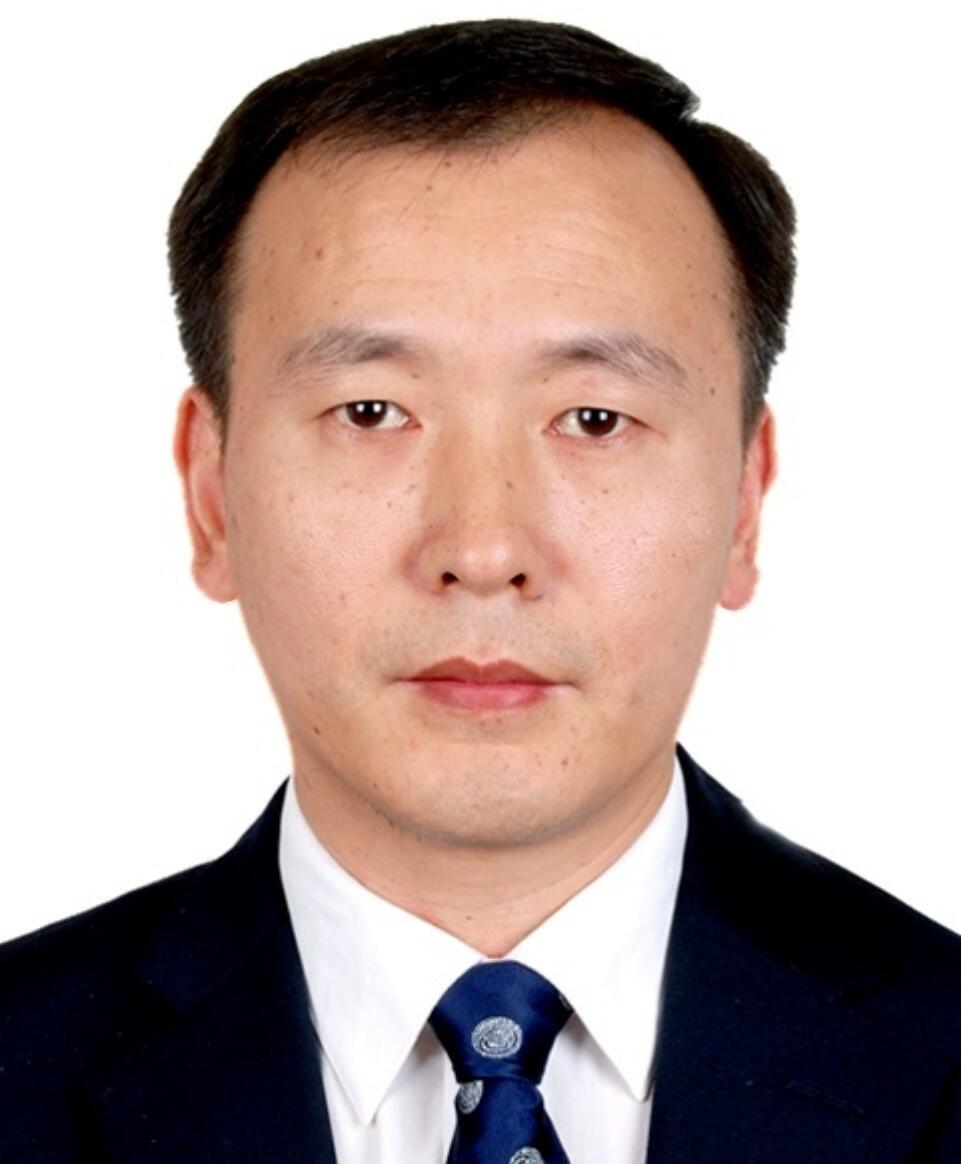}}]{Xiqi Gao} (xqgao@seu.edu.cn) received the Ph.D. degree in Electrical Engineering from Southeast University, China, in 1997. He joined the Department of Radio Engineering, Southeast University, in Apr. 1992 and became a professor since May 2001. He received the Science and Technology Awards of the State Education Ministry of China in 1998, 2006, and 2009, the National Technological Invention Award of China in 2011, and the 2011 IEEE Communications
Society Stephen O. Rice Prize Paper Award in the Field of Communications Theory. His current research interests include broadband multicarrier communications, MIMO wireless communications, channel estimation, and turbo equalization, and multirate signal processing for wireless communications. 
\end{IEEEbiography}

\begin{IEEEbiography}[{\includegraphics[width=1in,height=1.25in,clip,keepaspectratio]{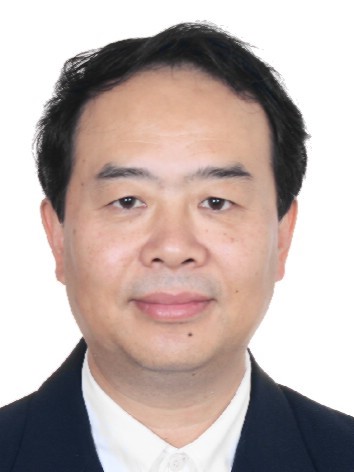}}]{Xiaohu You} (xhyu@seu.edu.cn) received his M.S. and Ph.D. degrees from Southeast University, Nanjing, China, in Electrical Engineering in 1985 and 1988, respectively. Since 1990, he has been working with National Mobile Communications Research Laboratory at Southeast University, where he is currently the director of the Lab. He has contributed over 100 IEEE journal papers and 3 books in the areas of adaptive signal processing, neural networks and their applications to communication systems. Now he is Secretary General of the FuTURE Forum, vice Chair of China IMT-2020 Promotion Group, vice Chair of China National Mega Project on New Generation Mobile Network. He was the recipient of the National 1st Class Invention Prize in 2011, and he was selected as IEEE Fellow in same year.
\end{IEEEbiography}

\begin{IEEEbiography}[{\includegraphics[width=1in,height=1.25in,clip,keepaspectratio]{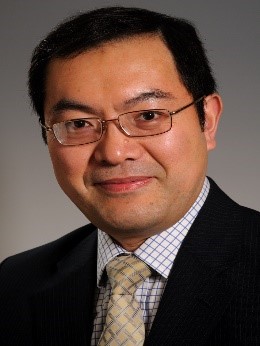}}]{Yang Hao} (y.hao@qmul.ac.uk) received the Ph.D. degree in Computational Electromagnetics from the
Center for Communications Research, University of Bristol, Bristol, U.K., in 1998. He was a Post-Doctoral Research Fellow with the School of Electronic, Electrical and Computer Engineering, University of Birmingham, Birmingham, U.K. He is currently a Professor of antennas and electromagnetics with the Antenna Engineering Group, Queen Mary University of London, London, U.K. His current research interests include computational electromagnetics, microwave metamaterials,
graphene and nanomicrowaves, antennas and radio propagation for body centric wireless networks, active antennas for millimeter/submillimeter applications, and photonic integrated antennas.
\end{IEEEbiography}

\end{document}